\documentclass{emulateapj}
\newcommand{\vecg}{\mbox{\boldmath $g$} {}}
\newcommand{\na}{\mbox{\boldmath $\nabla$}{}}

\begin{document}
\title{Is Holmberg-II beyond MOND theory?}
\author{F.~J.~S\'{a}nchez-Salcedo$^{1}$ and A.~M.~Hidalgo-G\'amez$^{2}$}
\affil{$^{1}$Instituto de Astronom\'\i a, Universidad Nacional Aut\'onoma
de M\'exico, \\
P.O. Box 70-264, Ciudad Universitaria, C.P. 04510, Mexico City, Mexico}

\affil{$^{2}$Departamento de F\'{\i}sica, Escuela Superior de F\'{\i}sica
y Matem\'aticas, IPN, \\ U.P. Adolfo L\'opez Mateos,
C.P. 07738, Mexico City, Mexico}

%\email{jsanchez@astroscu.unam.mx}

%\slugcomment{To be submitted to }
%\email{jsanchez@astroscu.unam.mx}

\begin{abstract}
We compare the basic properties and kinematics of two gas-rich dwarf galaxies:
KK246 and Holmberg II (HoII). 
HoII is $20$ times more luminous in the blue-band than KK246 and
its H\,{\sc i} mass is a factor of $6$ higher than in KK246.
However, the amplitudes of the rotation curves (at the last
measured point) of both galaxies are very similar,
of about $40$ km s$^{-1}$ at a galactocentric radius of $7$ kpc.
This fact is challenging for modified theories of gravity that predict
a one-to-one relation between gravity at any radius and the
enclosed baryonic mass in galaxies. In particular, MOdified
Newtonian Dynamics (MOND) predicts an asymptotic flat
velocity of $60$ km s$^{-1}$ in HoII. Since the baryonic mass of HoII is dominated
by the gas component, MOND overboosts its rotation speed even if the mass
of the stellar disk is taken negligibly small.
We conclude the rotation curve of HoII is probably inconsistent with MOND, unless
the inclination and distance are both fine-tuned to very unlikely values.

\end{abstract}


\section{Introduction}
The dynamical mass-to-light ratios $M/L$ derived in galaxies are usually
larger than the expected mass-to-light ratios of the stellar component,
indicating that they must contain dark matter.
Alternatively, one could argue that the discrepancy between total
mass and baryonic mass could tell us that the Newtonian law of
gravity is not governing the dynamics. In particular,
the MOdified Newtonian Dynamics (MOND) proposed
by Milgrom (1983) has been proven to be successful to reproduce
the kinematics of a significant fraction of
spiral galaxies without any dark matter (see Sanders \&
McGaugh 2002 for a review; Milgrom \& Sanders 2007; Sanders \& Noordermeer 2007;
Gentile et al.~2007). Swaters et al.~(2010) report
that the rotation curves of a quarter of their sample of $27$ dwarf 
and LSB galaxies are not adequately explained by MOND. 
Gentile et al.~(2011) also find that $3$ out of $12$ high-quality
rotation curves from the H\,{\sc i} Nearby Galaxy Survey (THINGS) 
are poorly fitted with MOND. Galaxies with poor MOND fits include NGC 2841,
NGC 3198 (Bottema et al.~2002), M31 (Corbelli \& Salucci 2007),
M33 (S\'anchez-Salcedo \& Lora 2005;
Corbelli \& Salucci 2007), UGC 4173 (Swaters et al.~2010),
UGC 6787 and UGC 11852 (Sanders \& Noordermeer 2007).
However, this is not necessarily a problem for MOND
because of  the inevitable uncertainties in the inclination
and in the distance of the galaxies,
and  because of the presence of warps and non-circular motions.

Recently, Milgrom (2011) highlights the individual importance of
the dwarf irregular galaxy KK246 as a new test, and conclude that the amplitude of its
rotation curve is correctly predicted using the MOND
prescription. He argues that it is rather puzzling for the $\Lambda$CDM paradigm to explain
why galaxies select exactly the velocities predicted by MOND.
In this note we discuss 
a counter-example. Indeed, the mass in gas of the dwarf irregular galaxy Holmberg II
(HoII) is a factor of $6$  higher than the gas mass in KK246.
Since stars hardly contribute to the baryonic mass in these galaxies, 
an analogue analysis dictates
that the expected MOND amplitude of HoII rotation curve should be 
noticeably larger than the amplitude of KK246 rotation curve.
However, the rotation curves of HoII and KK246 look very similar. 
In this paper we compare the properties of these galaxies and discuss
the implications of these findings in more detail.

\section{KK246 and HoII: similarities and differences}
KK246 (also referred to as ESO 461-036) and HoII (or DDO 50)
are gas-rich dwarf galaxies with more mass in gas than
in stars. Hence, the uncertainty of the stellar mass-to-light ratio
in mass models is less relevant than in high surface
brightness galaxies.
In Table \ref{table:parameters}, we give some basic properties of HoII 
and KK246.
 It is worthwhile noting that KK246 resides within
the nearby Tully void, while HoII belongs to the M81 group.

Careful analyses of H\,{\sc i} observations of HoII have been 
carried out by two independent groups: Bureau \& Carignan (2002)
and Oh et al.~(2011). Bureau \& Carignan (2002)
were able to infer the rotation curve until a radius of $\sim 20$ kpc. 
However, as these authors clearly stressed, the rotation curve derived
is well defined for galactocentric radius $R<10$ kpc because, for larger radii,
velocities were only measured in the approaching
side. Bureau \& Carignan derived a dynamical mass-to-light ratio of $16$.  
Recently, the THINGS high-resolution
data have significantly reduce observational uncertainties at $R<7$ kpc 
(Oh et al.~2011).

\begin{table*}
\begin{minipage}{126mm}
\caption[]{Comparison of the relevant parameters of the two galaxies}
\vspace{0.01cm}

\begin{tabular}{c c c c c c c c c}\hline
{Name} & {$D$} & {$M_{B}$} & $L_{B}$ & {$\left<i\right>$} & {$M_{\rm gas}$} & {$M_{\rm
bar}$} & $V_{\rm rot}(7 \, {\rm kpc})$ & References    \\
{}&Mpc &  & $10^{8}L_{\odot}$ & deg & $10^{8}M_{\odot}$ & $10^{8}M_{\odot}$ &km s$^{-1}$    \\

\hline
KK246 & 7.83 & -13.69 & 0.46 & $65^{\circ}$ & 1.5 & 2.0 &41 & Kreckel et al.~(2011) \\
HoII & 3.4 & -16.87 & 8.7 & $49^{\circ}$ & 9.0  & 10.5 & 37 & Walter et al.~(2008)\\
& & & & & & & & Oh et al.~(2011)  \\
\hline
\end{tabular}
\medskip\\
Column (1): Name of the galaxy. Column (2): Distance. Column (3):
Absolute B magnitude. Column (4): Total blue-band luminosity.
Column (5): Average value of the inclination within a radius of $7$ kpc.
Column (6): Total mass in gas. Column (7): Total baryonic mass. Column (8): Rotation velocity at a
galactocentric distance of $7$ kpc. Column (9): References.
\label{table:parameters}
\end{minipage}
\end{table*}

On the other hand, the dwarf irregular galaxy
KK246 has been studied by Kreckel et al.~(2011) with
the VLA and EVLA. They discovered an extended H\,{\sc i} disk
and were able to measure its rotation curve until a galactocentric 
radius of $7$ kpc.
They have estimated a dynamical mass-to-light ratio of $89$ in KK246.

\begin{figure}
 \plotone{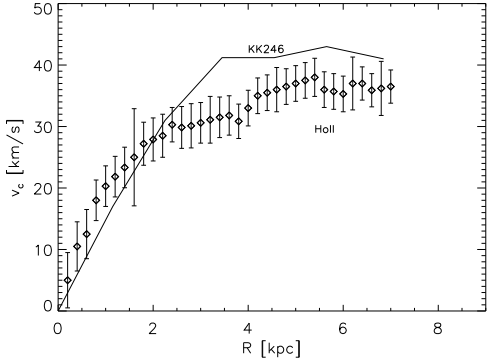}
 \caption{Measured rotation curves for KK246 (solid line) and HoII
(symbols).
}
 \label{fig:tworotationcurves}
 \end{figure}

In Figure \ref{fig:tworotationcurves}, we plot the observed rotation curves for
KK246 (Kreckel et al.~2011) and HoII (Oh et al.~2011). 
We see that both rotation curves are rather similar.  In the
case of HoII, the method of Oh et al.~(2011) minimizes the effect of
non-circular motions. The rotation curve of HoII was corrected for 
axisymmetric drift whereas this correction was not made for KK246. 
Axisymmetric drift corrections would cause a boost 
of a few km s$^{-1}$ in the circular velocities of KK246.

\section{The baryonic mass of HoII and KK246}
\label{sec:baryonic}
In MOND framework, the rotation velocity of an isolated galaxy is determined by its visible
(baryonic mass). 
For KK246, Kirby et al.~(2008) estimated a stellar mass of 
$5\times 10^{7}M_{\odot}$, which corresponds to stellar mass-to-light ratio
in the B-band of $\Upsilon_{\star}^{B}=1.1$
in solar units (see Table \ref{table:parameters}).
According to Kreckel et al.~(2011), the total gas mass of KK246
is $1.5\times 10^{8}M_{\odot}$ (assuming that the gas mass is $1.4$ times
the H\,{\sc i} mass). Therefore, the total baryonic
mass of KK246 is of $2\times 10^{8}M_{\odot}$, as used by Milgrom (2011). 

Using some relations derived from population synthesis models, 
Oh et al.~(2011) derived 
the stellar mass-to-light ratio in K-band, $\Upsilon_{\star}^{K}$,
for HoII.
With such a $\Upsilon_{\star}^{K}$-value (which we will refer to as the
nominal value), the stellar mass in the disk of HoII
is of $1.5\times 10^{8}M_{\odot}$. On the other hand,
Bureau \& Carignan (2002) found 
a H\,{\sc i} mass for this galaxy of $6.44\times 10^{8}M_{\odot}$, which corresponds to
a total mass in gas of $9\times 10^{8}M_{\odot}$.
Thus, the total (gas plus stars) baryonic mass in HoII is about
$10.5\times 10^{8}M_{\odot}$, which 
is a factor of $\sim 5$ larger than the baryonic mass in KK246.

\section{The rotation curve in MOND}
The Lagragian MOND field equations lead to a modified version of Poisson's equation
given by
\begin{equation}
\na\cdot \left[\mu\left(\frac{|\na\Phi|}{a_{0}}\right)\na\Phi\right] =4\pi G \rho,
\end{equation}
where $\rho$ is the density, $a_{0}$ is a universal acceleration of the order of 
$10^{-8}$ cm s$^{-2}$, and $\mu(x)$ is some interpolating function with
the property that $\mu(x)=x$ for $x\ll 1$ and $\mu(x)=1$ for $x\gg 1$ (Bekenstein \& Milgrom 1984).
To a good approximation, the {\it real} acceleration at
the midplane of a isolated, flattened axisymmetrical
system, $\vecg$, is related with
the Newtonian acceleration, $\vecg_{N}$, by:
\begin{equation}
\mu\left(\frac{|\vecg|}{a_{0}}\right)\vecg=\vecg_{N}.
\label{eq:algMOND}
\end{equation}
The two most popular choices for the interpolating function are
the ``simple'' $\mu$-function, suggested by Famaey \& Binney (2005),
\begin{equation}
\mu(x)=\frac{x}{1+x},
\end{equation}
and the ``standard'' $\mu$-function
\begin{equation}
\mu(x)=\frac{x}{\sqrt{1+x^{2}}},
\end{equation}
proposed by Milgrom (1983). For a sample of galaxies having a gradual transition
from the Newtonian limit in the inner regions to the MOND limit in the outer parts,
Famaey et al.~(2007) and Sanders \& Noordermeer (2007) conclude that the
plausability of the relative stellar mass-to-light ratios for bulge and disk, as well as
the generally smaller global $M/L$, lend support to the simple $\mu$-function
(see also Weijmans et al.~2008 and Gentile et al.~2011).

\begin{figure*}
 \plotone{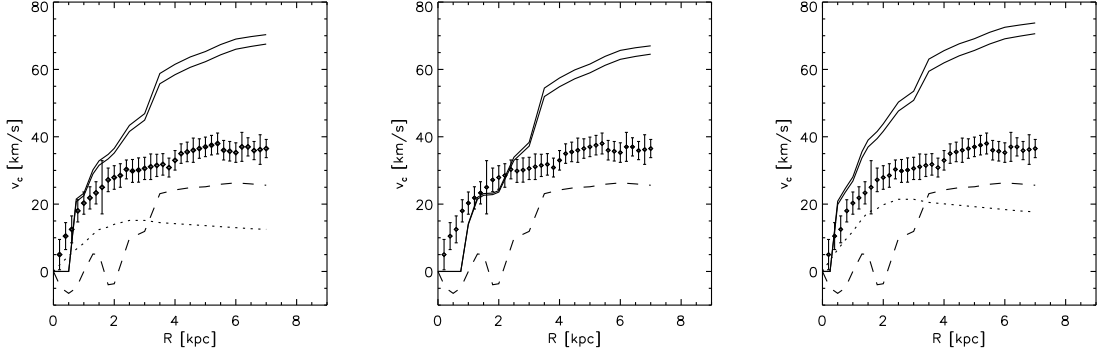}
 \caption{Rotation curve of HoII from Oh et al.~(2011)
together with the contributions of the stellar disk
(dotted lines) and gas (dashed lines) for the nominal
$\Upsilon_{\star}$-value (left panel), minimum disk plus gas
(central panel) and twice the nominal $\Upsilon_{\star}$-value.
The full lines represent the rotation curve according to MOND prescription
using the simple $\mu$-function (upper solid curves) or the standard 
$\mu$-function (lower solid curves). 
MOND is unable to provide the amplitude of the 
rotation curve. Here we take $a_{0}=1.2\times 10^{-8}$ cm s$^{-2}$.
}
 \label{fig:models_HoII}
 \end{figure*}

Certainly HoII is not at isolation; it is embbeded in the external 
gravitational field of M81 group. Since the modified Poisson equation
is nonlinear, the internal acceleration of a system depends on the 
external acceleration field $\vecg_{\rm ext}$ (Bekenstein \& Milgrom
1984). In order
to quantity the external field effect (EFE), it is important to compare 
the internal and external accelerations.
Assuming that the M81 group is bound, the external acceleration is
$0.7\times 10^{-10}$ cm s$^{-2}$ (Karachentsev et al.~2002).
The HoII internal accelerations are of $6\times 10^{-10}$ cm s$^{-2}$ and 
$2\times 10^{-10}$ cm s$^{-2}$ at $R=7$ kpc and $R=20$ kpc, respectively.
Thus, EFE should be small at $R<10$ kpc.
Indeed, the flatness of the rotation curve of HoII would be  
incompatible with HoII being dominated by the external field.

In the prescription of MOND, the asymptotic velocity 
is $(GMa_{0})^{1/4}$, where $M$ is the total baryonic mass of the
system. Consequently,
the asymptotic velocity in HoII is expected to be $50\%$ higher than
in KK246. In fact, using the estimates of the baryonic masses given in Section
\ref{sec:baryonic} and $a_{0}=1.2\times 10^{-8}$ cm s$^{-2}$,
MOND predicts correctly the asymptotic velocity ($42$ km s$^{-1}$) for KK246 but
overpredicts the HoII rotation speed ($63$ km s$^{-1}$). 
The predicted HoII rotational speed is in excess by $25$ km s$^{-1}$.

\begin{figure}
 \plotone{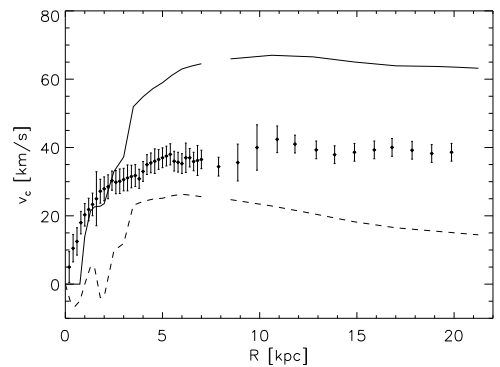}
 \caption{Combined rotation curve of HoII as measured by Oh et al.~(2011) 
(at $R<7$ kpc) and by Bureau \& Carignan (2002) at $R>7$ kpc.  
The dashed line represents the contribution to the
rotation curve of the gas disk.  The solid line represents the 
MONDian rotation curve in the ``minimum disk$+$gas'' assumption,
using the standard $\mu$-function and $a_{0}=1.2\times 10^{-8}$ cm s$^{-2}$.
All the variables have been rescaled for the adopted distance of
$D=3.4$ Mpc.
}
 \label{fig:hybrid}
 \end{figure}

Figure \ref{fig:models_HoII} shows the predicted HoII rotation curve in MOND under
various
$\Upsilon_{\star}^{K}$ assumptions [the nominal value as derived
by Oh et al.~(2011), ``minimum disk plus gas'', and twice the nominal
value]. The ``minimum disk plus gas'' includes the gas component and uses
the minimum value of $\Upsilon_{\star}$ compatible with the requirement that the theoretical
circular velocity must be positive  and reasonably smooth (i.e.~without unrealistic
steep gradients within $R<2$ kpc).
We see that the discrepancy between the observed and the predicted rotation
curves is very large. The effect of varying $\Upsilon_{\star}$ or the
interpolating function on the MOND circular velocity at the outer disk
is small. For illustration, Figure \ref{fig:hybrid} shows the combined rotation curve from 
Bureau \& Carignan (2002)
and Oh et al.~(2011) for the updated HoII distance of $3.4$ Mpc. 
The shift between the observed and the theoretical rotation curves is apparent.

Using a smaller value for $a_{0}$ would help to alleviate
the discrepancy. 
For instance, for a sample of $12$ galaxies,
Begeman et al.~(1991) found $a_{0}=(1.35\pm 0.51)\times 10^{-8}$ cm s$^{-2}$.
Gentile et al.~(2010) obtained $a_{0}=(1.22\pm 0.33)\times 10^{-8}$ cm s$^{-2}$
also for a sample of $12$ galaxies.
However, the MOND circular rotation speed at $7$ kpc for 
a value of $a_{0}$ at the lower end of the
best-fit interval, $a_{0}=0.9\times 10^{-8}$ cm s$^{-2}$,
is only a few km s$^{-1}$ slower. 

\begin{figure}
 \plotone{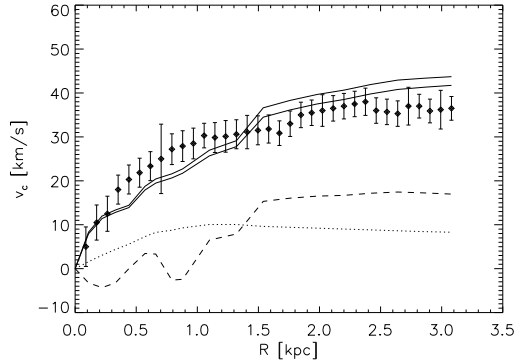}
 \caption{HoII rotation curve in MOND 
adopting the standard $\mu$-function with $a_{0}=0.9\times
10^{-8}$ cm s$^{-2}$ and a distance to the galaxy of $1.5$ Mpc, a factor $2.3$
closer than the nominal distance.
The key to lines is the same as in Fig.~\ref{fig:models_HoII}.  
}
 \label{fig:closer}
 \end{figure}

In the MOND prescription, 
the amplitude of the rotation curve can be accounted for 
by adopting a distance to HoII of $1.5$ Mpc and
$a_{0}=0.9\times 10^{-8}$ cm s$^{-2}$ (see Figure \ref{fig:closer}).
Given that the uncertainty in the distance is of $0.4$ Mpc (Karachentsev
et al.~2002), this likely
indicates that MOND cannot be made compatible with that rotation curve
by a reasonable adjustment of galaxy's distance.

A more delicate issue is the error resulting from the uncertainty in the inclination of the galaxy.
It turns out that if the inclination is taken as a free parameter,
a mean inclination of $25^{\circ}$ would yield a circular velocity
of $\sim 60$ km s$^{-1}$ at $R=7$ kpc. Since the uncertainty in the
inclination for this galaxy is about $10^{\circ}$ (Oh et al.~2011), 
the MOND rotation curve would be consistent with the observed curve only if the inclination 
is fine-tuned to very unlikely values. For rings at $R>12$ kpc,
Bureau \& Carignan (2002) required inclinations of $i=84^{\circ}$.
Even in the worst scenario
that the uncertainty in the inclination is of $30^{\circ}$, the
rotation speed at $R>12$ kpc would increase by a
factor of $\sin(84^{\circ})/\sin(54^{\circ})=1.25$ (from $\sim 40$ 
km s$^{-1}$ to $\sim 50$ km s$^{-1}$), which is not enough to
reconcile MOND with observations (see Fig.~\ref{fig:hybrid}).

\section{Conclusions}
van der Kruit (1995) noticed that the galaxies NGC 891 and NGC 7814
present very similar H\,{\sc i} kinematics but remarkably different light
distributions and claimed that this observation weakens the
appeal of MOND. Here we present two galaxies that rotate
at similar velocities but their baryonic masses  differ by a factor
of $5$. We have shown that the rotation curve of HoII is not
compatible with MOND unless the inclination is fine-tuned to
a very unlikely value. This inclination is the one required for
HoII to satisfy the baryonic Tully-Fisher relation.

\acknowledgements
We would like to thank Elias Brinks for his encouragement to make
these results public. This work was supported by CONACyT project
60526F and PAPIIT project IN121609.

{}


\begin{thebibliography}{}
\bibitem[Begeman et al.(1991)]{beg91}
Begeman, K.~G., Broeils, A.~H., Sanders, R.~H. 1991, MNRAS, 249,523
\bibitem[Bekenstein \& Milgrom(1984)]{bek84}
Bekenstein, J., \&  Milgrom, M. 1984, \apj, 286, 7
\bibitem[Bottema et al.(2002)]{bot02}
Bottema, R., Pesta\~{n}a, J. L. G., Rothberg, B., Sanders, R. H. 2002, A\&A,
393, 453
\bibitem[Bureau \& Carignan(2002)]{car02}
Bureau, M., \& Carignan, C. 2002, \aj, 123, 1316
\bibitem[Corbelli \& Salucci(2007)]{cor07}
Corbelli, E., \& Salucci, P. 2007, \mnras, 374, 1051
\bibitem[Famaey \& Binney(2005)]{fam05}
Famaey, B., \& Binney, J. 2005, \mnras, 363, 603
%\bibitem[Gentile et al.(2007)]{gen07}
%Gentile, G., Famaey, B., Combes, F., Kroupa, P., Zhao, H. S., \& Tiret, O.
\bibitem[Gentile et al.(2011)]{gen11}
Gentile, G., Famaey, B., \& de Blok, W. J. G. 2011, \aap, 527, 76
2007, \aap, 472, L25
\bibitem[Karachentsev et al.(2002)]{kar02}
Karachentsev, I. D. et al. 2002, \aap, 383, 125
%\bibitem[Huchra \& Geller(1982)]{huc82}
%Huchra, J. P., \& Geller, M. J. 1982, \apj, 257, 423
\bibitem[Kirby et al.(2008)]{kir08}
Kirby, E. M., Jerjen, H., Ryder, S. D., \& Driver, S. P. 2008, \aj, 136, 1866
\bibitem[Kreckel et al.(2011)]{kre11}
Kreckel, K., Peebles, P. J. E., van Gorkom, J. H., van de Weygaert, R.,
\& van der Hulst, J. M. 2011, arXiv:1103.5798
\bibitem[Milgrom(1983)]{mil83}
Milgrom, M. 1983, ApJ, 270, 365
\bibitem[Milgrom(2011)]{mil11}
Milgrom, M. 2011, arXiv:1104.1118 
\bibitem[Milgrom \& Sanders(2007)]{mil07}
Milgrom, M., \& Sanders, R. H. 2007, \apj, 658, L17
\bibitem[Oh et al.(2011)]{oh11}
Oh, S.-H., de Blok, W. J. G., Brinks, E., Walter, F., \& Kennicutt, R. C. 2011, in press (arXiv:1011.0899)
\bibitem[S\'anchez-Salcedo \& Lora(2005)]{san05}
S\'anchez-Salcedo, F. J., Lora, V. 2005, in Progress in Dark Matter
Research, ed.~J.~Val Blain (Nova Publications: New York)
\bibitem[Sanders(1996)]{san96}
Sanders, R. H. 1996, ApJ, 473, 117
\bibitem[Sanders \& McGaugh(2002)]{san02}
Sanders, R. H., McGaugh, S. S. 2002, ARA\&A, 40, 263
\bibitem[Sanders \& Noordermeer(2007)]{san07}
Sanders, R. H., \& Noordermeer, E. 2007, \mnras, 379, 702
\bibitem[Swaters et al.(2010)]{swa10}
Swaters, R. A., Sanders, R. H., \& McGaugh, S. S. 2010, \apj, 718, 380
\bibitem[van der Kruit(1995)]{kru95}
van der Kruit, P. C. 1995, in Stellar Populations, IAU Symp.~164, 
Editors P. C. van der Kruit, G. Gilmore, 
Kluwer Academic Publishers, Dordrecht, p.~205
\bibitem[Walter et al.(2008)]{wal08}
Walter, F., Brinks, E., de Blok, W. J. G., Bigiel, F., Kennicutt,
R. C., Thornley, M., \& Leroy, A. 2008, \aj, 136, 2563
\bibitem[Weijmans et al.(2008)]{wei08}
Weijmans, A.-M., Krajnovic, D., van de Ven, G., Oosterloo, T. A., Morganti, R., \& de Zeeuw, P. T.
2008, \mnras, 379, 1343
\end{thebibliography}
\end{document}